\documentclass[a4paper,fleqn,usenatbib]{mnras}

\usepackage{mathptmx}
\usepackage[T1]{fontenc}
\usepackage{ae,aecompl}

\usepackage{graphicx}	% Including figure files
\usepackage{amsmath}	% Advanced maths commands
\usepackage{amssymb}	% Extra maths symbols

\newcommand{\fermi}{{\it Fermi}-LAT}

\title[Optical spectroscopy of {\it Fermi}-LAT BCUs]{Optical spectroscopic classification of a selection of Southern Hemisphere {\it Fermi}-LAT unclassified blazars\thanks{Based, on observations made with the Southern African Large Telescope (SALT) under programs 2014-2-SCI-055 and 2015-1-SCI-053 (PI L. Klindt).}}

\author[L. Klindt et al.]{
L. Klindt,$^{1}$
 B. van Soelen,$^{1}$\thanks{E-mail: vansoelenb@ufs.ac.za}
P.J. Meintjes,$^{1}$
and P. V\"ais\"anen$^{2,3}$
\\
$^{1}$Department
of Physics, University of the Free State, 9300, Bloemfontein, South Africa\\
$^2$South African Astronomical Observatory, PO Box 9, Observatory 7935, Cape Town, South Africa\\
$^3$Southern African Large Telescope, PO Box 9, Observatory 7935, Cape Town, South Africa\\
}

\date{Accepted XXX. Received YYY; in original form ZZZ}

\pubyear{2016}

\begin{document}
\label{firstpage}
\pagerange{\pageref{firstpage}--\pageref{lastpage}}
\maketitle

% Abstract of the paper
\begin{abstract}
% This is a simple template for authors to write new MNRAS papers.
% The abstract should briefly describe the aims, methods, and main results of the paper.
% It should be a single paragraph not more than 250 words (200 words for Letters).
% No references should appear in the abstract.
The {\it Fermi}-LAT has detected more than 3000 sources in the GeV $\gamma$-ray regime. The majority are extra-galactic and these sources are dominated by blazars. However, $\sim28$ per cent of the sources in {\it Fermi} 3LAC are listed as blazar candidates of uncertain type (BCU). Increasing the number of classified {\it Fermi}-LAT sources is important for improving our understanding of extra-galactic $\gamma$-ray sources and can be used to search for new very high energy sources.  We report on the optical spectroscopy of seven selected unclassified BCU sources during 2014 and 2015 undertaken using the SAAO 1.9-m and Southern African Large Telescope (SALT). Based on the identified spectral lines we have classified three of the sources as FSRQs and the remaining four as BL Lac objects, determining the redshift for four sources. 
\end{abstract}

% Select between one and six entries from the list of approved keywords.
% Don't make up new ones.
\begin{keywords}
BL Lacertae objects: general -- galaxies: active  -- gamma rays: galaxies -- quasars: general
\end{keywords}

%%%%%%%%%%%%%%%%%%%%%%%%%%%%%%%%%%%%%%%%%%%%%%%%%%

%%%%%%%%%%%%%%%%% BODY OF PAPER %%%%%%%%%%%%%%%%%%

\section{Introduction}

The Large Area Telescope (LAT) on-board the {\it Fermi} $\gamma$-ray space telescope has been in operation since 2008 and provides an all sky detection of $\gamma$-ray sources in, primarily, the MeV to GeV range (though with the new pass 8 analysis, detections are possible up to $\sim1$~TeV). The majority of these sources are extra-galactic and $\sim59$ per cent of all sources within the 3$^\rmn{rd}$ \fermi\ source catalogue \citep[3FGL,][]{acero15} are classified as Active Galactic Nuclei (AGN) or AGN candidates.  The third and most recent \fermi\ catalogue of AGN \citep[3LAC,][]{ackermann15} lists 1444 sources (in the clean sample) above a galactic latitude of $|b|>10$\degr\ which are associated with AGN. 

AGN are powered by the accretion of material onto a central supermassive blackhole (SMBH). In radio loud sources, this is accompanied by the presence of a relativistic jet. The observed properties of these sources are strongly dependent on the viewing angle and when our line of sight lies close to the direction of jet propagation the observed emission is dominated by Doppler boosted non-thermal emission originating from the jet; these sources are collectively known as blazars \citep[e.g.][]{urry95}.  The emission from blazars is known to show rapid variability across the broad wavelength range (from radio to TeV $\gamma$ rays), as well as variable polarization detected in radio and optical. 

\fermi\ AGN are dominated by blazar sources, ($\sim$70~per cent), while there are only 24 AGN of other types ($\sim 2$ per cent). The remaining 402 sources ($\sim28$ per cent) are classified as blazar candidates of uncertain type (BCU). These are sources that are associated with extra-galactic counterparts, show blazar characteristics, but lack reliable spectral classification.  Blazars are sub-divided into BL Lacs or Flat Spectrum Radio Quasars (FSRQs) based on the observed strength of the optical spectral lines \citep{marcha96, landt04}.   In addition to the unclassified sources, $\sim36$ per cent of all \fermi\ detected sources remain unidentified \citep{acero15}. 

The Spectral Energy Distributions (SEDs) of blazars show two clear components and blazars are also classified depending on where the lower energy component peaks ($\nu_{\rm peak}$). The lower energy component (radio to UV/X-ray)  is produced by electron synchrotron radiation while the higher energy component (UV/X-ray to $\gamma$-ray energies) can be reproduced through either inverse Compton scattering of synchrotron or external photons, or through hadronic processes \citep[e.g.][]{boettcher13, petropoulou15}.   
The SED based sub-divisions are low--, intermediate--, or high-synchrotron peaked (LSP, ISP, or HSP) sources, with peak frequencies of $\nu_{\rm peak} < 10^{14}$~Hz, $10^{14}~{\rm Hz} \leq \nu_{\rm peak} < 10^{15}$~Hz, or $\nu_{\rm peak} \geq 10^{15}$~Hz, respectively \citep[e.g.][]{padovani95, ackermann11}.

In addition to the SED classification, the $\gamma$-ray observations also show an indication that the FSRQs and BL Lacs follow different distributions. For example, the distribution in $\gamma$-ray photon index, $\Gamma$, for classified sources shows that FSRQs are softer ($\langle \Gamma \rangle =2.44\pm0.20$) than the BL Lacs ($\langle \Gamma \rangle = 2.01\pm0.25$) and form two groups \citep[see e.g. fig.~7 in][]{ackermann15}.  However, the distribution is broad and this alone cannot be used to classify the sources.

Increasing the number of classified sources improves our understanding of the origin of extra-galactic $\gamma$-ray emission and optical classifications have been undertaken by, for example, \citet{shaw09} and \citet{alvarezc_respo16}. The optical spectroscopy also provides a method to determine the redshift of these sources. In addition to these observational campaigns, machine learning techniques are also being developed to classify sources based purely on their $\gamma$-ray properties \citep[e.g][]{hassan13,chiaro16}. However, the final classification depends on the optical spectral properties.

Supplementary to this, the 2LAC BCU sources provide an important resource to identify new very high energy (VHE) emitting $\gamma$-ray candidates that may be observable with ground based atmospheric Cherenkov telescopes. Observations at VHE are, however, constrained by $\gamma\gamma$ absorption due to the interaction of $\gamma$ rays with the extragalactic background light (EBL) and are limited to $z\lesssim1$ \citep[e.g.][]{gould66,hauser01,aharonian06}. 
The identification of new very high energy candidate sources at low redshifts is an important outcome from such observations.

Here we report on optical spectroscopy undertaken of a selection of 7 unclassified BCU sources during 2014 and 2015. The paper is structured as follows: section~\ref{sec:classification} summarises the classification applied to the sources, section~\ref{sec:source} discusses the selection of sources, section~\ref{sec:obs} presents the observations and data analysis. The results are presented in section~\ref{sec:results} and the discussion and conclusions in section~\ref{sec:conclusion}. 

\section{Source classification criteria}
\label{sec:classification}

The sub-division of blazars depends on the observed strength of their optical spectral lines and BL Lacs show no or weak emission lines while FSRQs exhibit both strong narrow and broad emission lines. Blazars are classified as BL Lacs if they show featureless spectra or exhibit emission lines with equivalent widths $|W_\lambda|\leq 5$~\AA{} while FSRQs show emission lines with widths $|W_\lambda| > 5$~\AA{} \citep[e.g.][]{landt04, marcha96,stickel91, stocke91}.  In this study we will use this as the dividing line between these two classes.

When strong emission lines are present, these originate from clouds in the Broad Line Region (BLR) and the Narrow Line Region (NLR) which surround the SMBH. The BLR lies close to the SMBH at a distance of $5-50\times10^{-2}$~pc, consists of fast-moving clouds (typical velocities of $\sim3000$~km~s$^{-1}$) producing broadened lines with typical widths on the order of 10--100~\AA{}. The NLR lies further out ($10^2 - 10^4$~pc), moving at lower speeds (200--700~km~s$^{-1}$), producing narrower emission lines. 

For BL Lac sources, with no emission lines, a characteristic feature is a weak Ca break. The Ca H\&K break is due to stellar absorption present in the spectra of elliptical galaxies. 
The strength of the Ca~H\&K break is measured by, $K_{4000} = (f^{+} - f^{-})/f^{+}$~
where $f^{-}$ and  $f^{+} $ are the average flux values between  3750--3950~\AA{}, and 4050--4250~\AA{}, respectively (in the rest frame of the galaxy).  The strength of the Ca~H\&K break is typically K$_{4000}$ $\sim$ 50 per cent in non-active elliptical galaxies \citep{dressler87} but for BL Lacs this feature is diluted (K$_{4000}$ $\la$ 40 per cent) due to the non-thermal emission from the jet \citep[e.g.][]{landt02}.

\section{Source selection}
\label{sec:source}

When observations began, the unclassified sources were selected from the 2LAC \citep{ackermann11} but all sources remain unclassified in the 3LAC. All selected sources are significantly detected by {\it Fermi} with a test statistic > 25. Here we report on seven sources from our sample; for the other sources, the objects have yet to be observed, or the current signal to noise ratio is currently too low. 
The counterparts that we have chosen are those identified within the {\it Fermi}-LAT catalogues as the most probable sources. These identifications are made using three statistical association methods, namely a Bayesian association method, a Likelihood Ratio association method and a $\log N - \log S$ association method \citep[see][for detailed discussions]{ackermann11,ackermann15}. 
The sources were selected taking the following properties into account:
\begin{description}
 \item[\it High galactic latitudes:] we only selected sources above $|b|>10$\degr\ from 2LAC to eliminate contamination from the galactic sources.
 \item[\it Counterparts - the 95\% error radius:] We ensured that the potential optical counterparts of the candidate AGN given were within the 95\% error radius reported in 3LAC. 
 \item[\it Observability:] For this initial project we restricted our search to Southern Hemisphere sources that are observable with South African based telescopes.
\item[\it Redshift measurements:] We selected sources that either lacked spectra or had spectra with signal to noise ratios (S/N) that were too low for a confident classification or redshift measurement from previous observations.
\end{description}

The selected \fermi\ targets with their identified counterparts are listed in Table~\ref{tab:sources} and the $\gamma$-ray properties are given in Table~\ref{tab:fermi_lat_prop}. 

\begin{table*}
	\centering
	\caption{The radio and optical counterparts for the selected BCU targets.}
	\label{tab:sources}
	\begin{tabular}{lllll} \hline 
3FGL name  &  2FGL name  &  2LAC counterpart   &  Radio counterpart   &  Optical counterpart   \\ \hline
J0045.2-3704  &  J0044.7-3702  &  PKS 0042-373   &   PKS J0045-3705   &  NVSS J004512.0-370549   \\
J0200.9-6635  &  J0201.5-6626  &  PMN J0201-6638  &  PMN J0201-6638   &  PMN J0201-6638 ID   \\
J0644.3-6713  &  J0644.2-6713  &  PKS 0644-671   &  PKS 0644-671   &  PKS 0644-671 ID  \\
J0730.5-6606  &  J0730.6-6607  &  PMN J0730-6602   &  PMN J0730-6602   &  QORG J073049.6-660219  \\
J1218.8-4827  &  J1218.8-4827  &  PMN J1219-4826   &  PMN J1219-4826   &  QORG J121902.2-482628  \\
J1407.7-4256  &  J1407.5-4257  &  CGRaBS J1407-4302   &  CGRaBS J1407-4302   &  QORG J140739.8-430234  \\
J2049.7+1002  &  J2049.8+1001  &  PKS 2047+098    &  PKS 2047+098   &  CGRaBS J2049+1003  \\ \hline
	\end{tabular}
\end{table*}

\begin{table}
\centering
\caption{Gamma-ray properties of the sources measured in the {\it Fermi}-LAT 3FGL catalogue \citep{ackermann15}. The table lists the photon flux, the photon index $\Gamma$ (0.1--100~GeV) and the significance of the detection, $\sigma$ (0.1--300~GeV).} 
\label{tab:fermi_lat_prop}

\begin{center}
\begin{tabular}{lccc} \hline
3FGL name & Flux ($1-100$~GeV) & $\Gamma$ & $\sigma$ \\
& $(\times10^{-10}$~cm$^{-2}$~s$^{-1})$  \\ \hline
J0045.2-3704 & $ 5.6 \pm 0.8 $ & $ 2.54 \pm 0.08 $ & 13.6 \\
J0200.9-6635 & $ 3.4 \pm 0.6 $ & $ 2.33 \pm 0.13 $ & 6.9 \\
J0644.3-6713 & $ 39.5 \pm 1.6 $ & $ 2.1 \pm 0.03 $ & 47.8 \\
J0730.5-6606 & $ 4.9 \pm 0.9 $ & $ 1.79 \pm 0.13 $ & 8.6 \\
J1218.8-4827 & $ 6.1 \pm 0.9 $ & $ 2.34 \pm 0.11 $ & 8.1 \\
J1407.7-4256 & $ 5.9 \pm 0.9 $ & $ 2.14 \pm 0.13 $ & 8.2 \\
J2049.7+1002 & $ 8.3 \pm 1 $ & $ 2.4 \pm 0.08 $ & 11.1 \\ \hline
\end{tabular}
\end{center}

\end{table}

\section{Observations}
\label{sec:obs}

Spectroscopic observations were undertaken using the SAAO 1.9-m and Southern African Large Telescope \citep[SALT;][]{odonoghue06}, both of which are located at the South African Astronomical Observatory (SAAO). 

\subsection{SAAO 1.9-m telescope}

Sources were observed during a two week period between 21 May - 3 June 2014 using the grating spectrograph on the SAAO 1.9-m telescope (ID 2013.12.Q2-03). All observations were performed with grating 7 at a grating angle of 17.1\degr, with a 1.5\arcsec\ slit width, providing a wavelength range of 3700--7900~\AA{} at a resolution of $\sim5$~\AA{}. Wavelength calibration was performed using the standard CuAr arc lamp. The mean seeing during the SAAO 1.9-m observations was 1.5\arcsec.

\subsection{SALT/RSS}
\label{sec:salt_rss}

Spectroscopic observations were undertaken with SALT using the RSS spectrograph \citep{burgh03} during two observing semesters. During the first semester (2014 Semester II) we observed two brighter targets (V < 18 mag) and during the second semester (2015 Semester I) we observed four fainter targets (V $\sim 20$~mag). All observations undertaken during the 2014 Semester II, were obtained in clear, grey (lunar illumination 15\% -- 85\%) conditions, whereas the observations for the fainter targets during the 2015 Semester~I were performed in clear dark (lunar illumination < 15\%) conditions. The average seeing for the observations was 1.8\arcsec.

The RSS detector consists of three mosaicked chips, which results in two small gaps in the wavelength dispersion direction. In order to avoid possibly missing lines due to this gap, observations were done in two camera (grating) angles.  For each observation two exposures were obtained. For the fainter sources,
a 10\arcsec\ dither along the slit between exposures was made to mitigate fringing effects in the red part of the spectrum.
The different instrument configurations, including wavelength ranges and spectral resolutions obtained, which we utilized for each semester are
shown in Table~\ref{tab:rss_conf}. The total exposure times for the targets are listed in Table~\ref{tab:exp_time}.

\begin{table*}
	\begin{center}
% 		\scriptsize
		\caption{\label{salt configurations} The RSS configurations utilized for the SALT 2014 Semester-II and 2015 Semester-I observations. Configurations A and C correspond to the `blue' configurations, whereas configurations B and D are the `red' configurations. Note that the wavelength range when using the lower resolution grating (pg0300) is constrained by the atmospheric cut-off in the blue and by instrumental efficiencies in the red.  } \label{tab:rss_conf}  
		\begin{tabular}[c]{cccccccc}
			\hline
			Configuration & Slit width & Grating & Grating angle & Wavelength range & Resolution & Camera binning & Arc lamp \\
			& (\arcsec) & & ($\degr$) & (\AA{}) & (\AA{}) & &  \\
			\hline
			\multicolumn{8}{c}{2014 Semester-II}\\
			\hline
			A & 0.6 & pg0300 & 5.0 & 3200 -- 9500 & $\sim$ 7.1 & 4$\times$2 & Ar \\
			B & 0.6 & pg0300 & 5.75 & 3200 -- 9500 & $\sim$ 7.1 & 4$\times$2  & Ne \\
			\hline
			\multicolumn{8}{c}{2015 Semester-1}\\
			\hline
			C & 2.0 & pg0900 & 14.75 & 4061 -- 7124 & $\sim$ 7.6 & 4$\times$2  & Ar \\
			D & 2.0 & pg0900 & 19.25 & 5751 -- 8741 & $\sim$ 7.4 & 4$\times$2 & Ne \\
			\hline
		\end{tabular}
	\end{center}
\end{table*}

\begin{table*}
 \begin{center}
\caption{Total exposure times  for all targets observed with the SAAO 1.9-m and SALT for the different RSS configurations. The V magnitudes are from the USNO-B1 catalogue as listed in the {\it Fermi}-2LAC. }
\label{tab:exp_time}
{%
\newcommand{\mc}[3]{\multicolumn{#1}{#2}{#3}}
\begin{center}
\begin{tabular}{lcccccc}\hline
3FGL name & V mag & SAAO 1.9-m &  \mc{4}{c}{SALT}\\
 &  & & A & B & C & D\\
 & &(s) & (s) & (s) & (s) & (s)\\ \hline
J0045.2-3704 &19.60 & - & - & - & 5000 & 5200\\
J0200.9-6635 &20.56& - & - & - & 2700 & 5400\\
J0644.3-6713 &20.69& - & - & - & 2600 & -\\
J0730.5-6606 &15.13& 9000 & 300 & 300 & - & -\\
J1218.8-4827 &17.53&  9600 & 568 & 458 & - & -\\
J1407.7-4256 &17.47& 14400 & - & - & 740 & 652\\
J2049.7+1002 &-& 9600 & - & - & - & - \\ \hline
\end{tabular}
\end{center}
}%

 \end{center}

\end{table*}

\subsection{Data reduction}

The data reduction was performed using the standard {\sc iraf}  tasks within the {\sc pyraf} package \citep{tody86}.  This included bias correction, flat-correction, background subtraction, and wavelength and flux calibration using the {\sc noao.imred.ccdred}, {\sc noao.onedspec} and {\sc noao.twodspec} packages. Cosmic ray removal was performed using the Laplacian Cosmic Ray Identification \citep{van_dokkum01}. 

The CCD pre-processing of the RSS observations was performed by the SALT {\sc pyraf} pipeline \citep{crawford10}. This includes bias and overscan subtraction, gain and cross-talk corrections, and detector mosaicking. 

Arc observations for wavelength calibration were performed before and after all science observations for the SAAO~1.9-m observations, and after science for all SALT observations. 

As mentioned above (section~\ref{sec:salt_rss}), some observations undertaken with the RSS included a dither to correct for fringing effects. OH emission lines cause fringing effects at longer wavelengths in the target image which are mostly corrected with background subtraction. However, since our targets are optically faint, we took two separate spectra shifted along the slit by approximately 10\arcsec, to remove the remainder of the fringes. After background correction has been performed, both of the two-dimensional target frames are subtracted from each other, creating two images, each with the target trace still visible, as well as a darkened trace due to the position of the other frame's trace.

Flux calibration was performed for all sources using the spectroscopic standards LTT 2415, LTT 4364, EG 21 and EG 273. However, corrections for the difference in slit width and the changing size of the SALT pupil have not been performed and, therefore, the fluxes do not reflect absolute values while the relative spectral shapes are accurate.

Finally, since the targets are faint, multiple observations per target were required and the source spectra were stacked to increase the signal-to-noise ratio of all observations.

The measurement of the identified spectral features was performed with {\sc spectool} within {\sc iraf}. Spectral features were identified for each target and the positions of the lines were identified from the centre of a Gaussian profile fitted to the spectral features. The equivalent widths, $W_\lambda$, of the spectral lines were determined from a Gaussian profile fitted to each line and the uncertainty was determined from \citep{vollmann06}
\[
 \sigma(W_\lambda) = \sqrt{1 + \left(\frac{\bar{F_\text{c}}}{\bar{F}}\right)} \left(\frac{\Delta \lambda - W_\lambda}{\text{S/N}}\right), 
\]
where $\bar{F_\text{c}}$ and $\bar{F}$ are the normalized average flux of the continuum and of the line at the wavelength $\lambda$, respectively, $\Delta \lambda$ is the wavelength interval, and S/N is the signal-to-noise ratio. 

Similarly the Full Width at Half Maximum (FWHM) obtained from the fit was corrected for instrumental broadening by
\[
 \text{Corrected width} = \sqrt{(\text{Uncorrected width})^2 - \delta^2},
\]
where $\delta$ is the wavelength resolution of the spectrograph.

\section{Results}
\label{sec:results}

The spectroscopic results are presented for the individual sources below. For all spectra, the shape corrected spectra are shown, along with the normalised spectra obtained by dividing by the continuum. For all spectra telluric and interstellar medium lines are indicated by $\oplus$ and IS, respectively. If spectral lines are identified these are used to determine the redshift of a source, where the statistical error reported is the standard deviation of the redshifts measured for each individual spectral line.

\subsection{3FGL J0045.2-3704}

\begin{figure}
	% To include a figure from a file named example.*
	% Allowable file formats are eps or ps if compiling using latex
	% or pdf, png, jpg if compiling using pdflatex
	\includegraphics[width=\columnwidth]{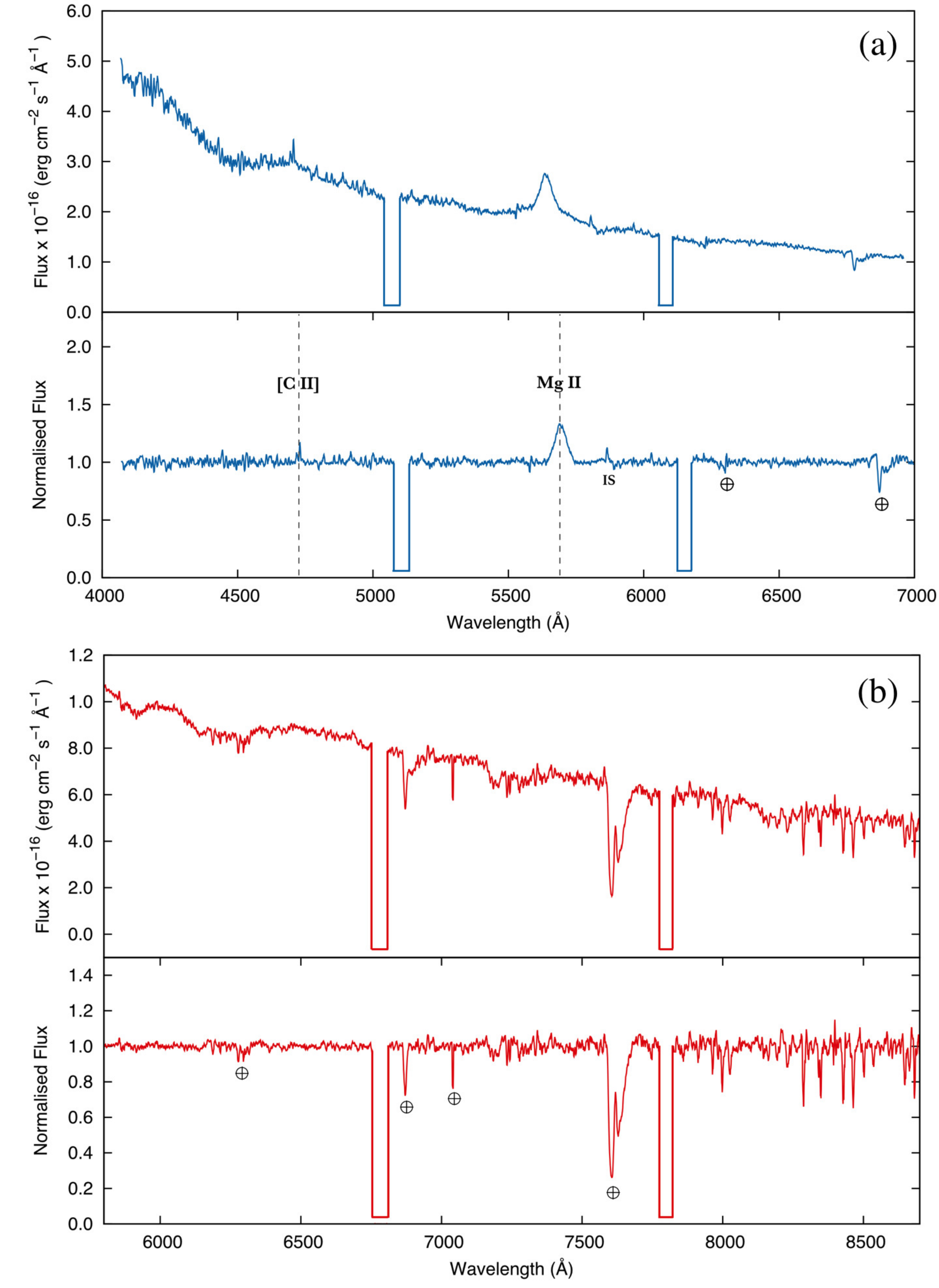}
    \caption{The average 3FGL J0045.2-3704 spectrum obtained with the RSS/SALT. (a) The average blue configuration spectrum. 
    The broad [C~{\sc ii}] and Mg~{\sc ii} lines are present. 
    (b) Average red configuration spectrum. No significant spectral features were detected. The gaps in the spectra are due to the gaps between the CCD chips in the RSS. 
}
    \label{fig:J0044}
\end{figure}

This source was observed twice with SALT using the C (blue) and D (red) RSS configurations, and the averaged spectrum is shown in Fig.~\ref{fig:J0044}. In the blue spectrum [C {\sc ii}] ($\lambda_0 = 2326.93$~\AA{}) and Mg~{\sc ii} ($\lambda_0  = 2797.99$~\AA{}) are detected at 4730~\AA{} and 5689~\AA{}. This places the source at a redshift of $z = 1.0331\pm 0.0004$. The weaker [C~{\sc ii}] feature has an equivalent width of $-1.5\pm0.8$~\AA{}, while the stronger Mg~{\sc ii} line is $-17.9\pm3.4$~\AA{} which classifies the source as a FSRQ. The [C {\sc ii}] and Mg {\sc ii} emission lines have FWHM values of 808.6~km~s$^{-1}$ and 2777.1~km~s$^{-1}$, respectively, suggesting that the Mg~{\sc ii} originates from the BLR while the [C {\sc ii}] line originates from the NLR.

\subsection{3FGL J0200.9-6635}

\begin{figure}
 	\includegraphics[width=\columnwidth]{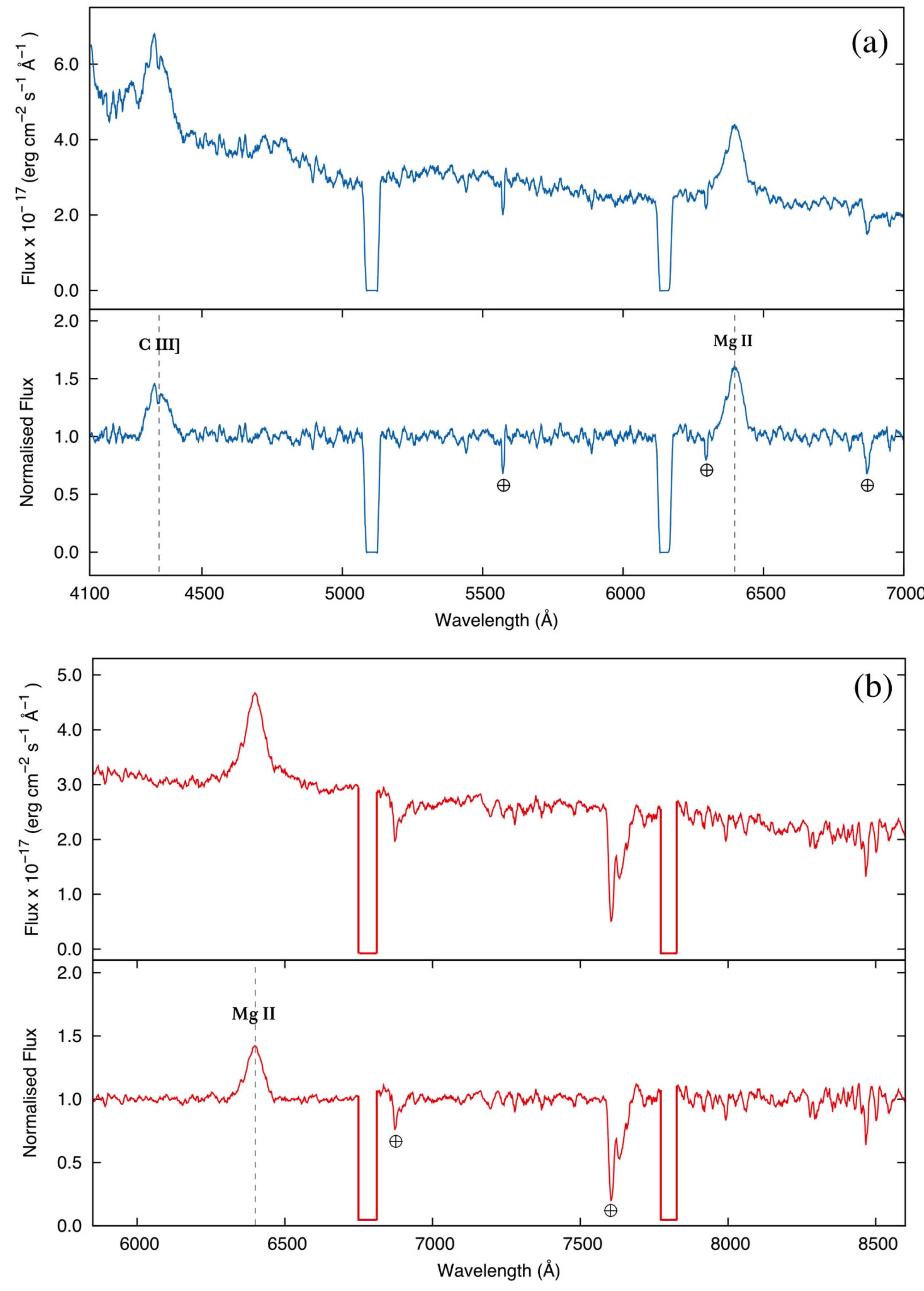}
	\caption{The average 3FGL J0200.9-6635 spectrum obtained with the RSS/SALT. (a) The blue configuration spectrum shows broad emission line features from C~{\sc iii}] and Mg~{\sc ii}. (b) Average red spectrum. The broad Mg~{\sc ii} emission line is detected}
	\label{fig:J0201}
\end{figure}

We obtained both blue (configuration C) and red (configuration D) RSS spectra for 3FGL J0200.9-6635 as shown in Fig.~\ref{fig:J0201}. Two emission lines, C~{\sc iii}] ($\lambda_{0,{\rm vac}} = 1908.73$~\AA{}) and Mg~{\sc ii} ($\lambda_0 = 2797.99$~\AA{}), are detected at $\sim4300$~\AA{} and $\sim6400$~\AA{}, respectively. This places the source at a redshift of $z=1.28\pm0.01$. The source is classified as an FSRQ due to the broad emission features with equivalent widths of $|W_{\lambda, \text{C{ {\sc III}}]}}| = 34\pm13$~\AA{} and $|W_{\lambda,\text{Mg{\sc ii}}}| = 42\pm12$~\AA{}, for the C~{\sc iii}] and Mg~{\sc ii} lines, respectively. This classifies the source as a FSRQ. Similarly, the broad emission lines (FWHM$_\text{C\,{\sc iii}}=5948.3$~km~s$^{-1}$, FWHM$_\text{Mg\,{\sc ii},blue}=2825.8$~km~s$^{-1}$ and FWHM$_\text{Mg\,{\sc ii},red}=2915.3$~km~s$^{-1}$) indicate the emission originates from the BLR.

\subsection{3FGL~J0644.3-6713}

\begin{figure}
 	\includegraphics[width=\columnwidth]{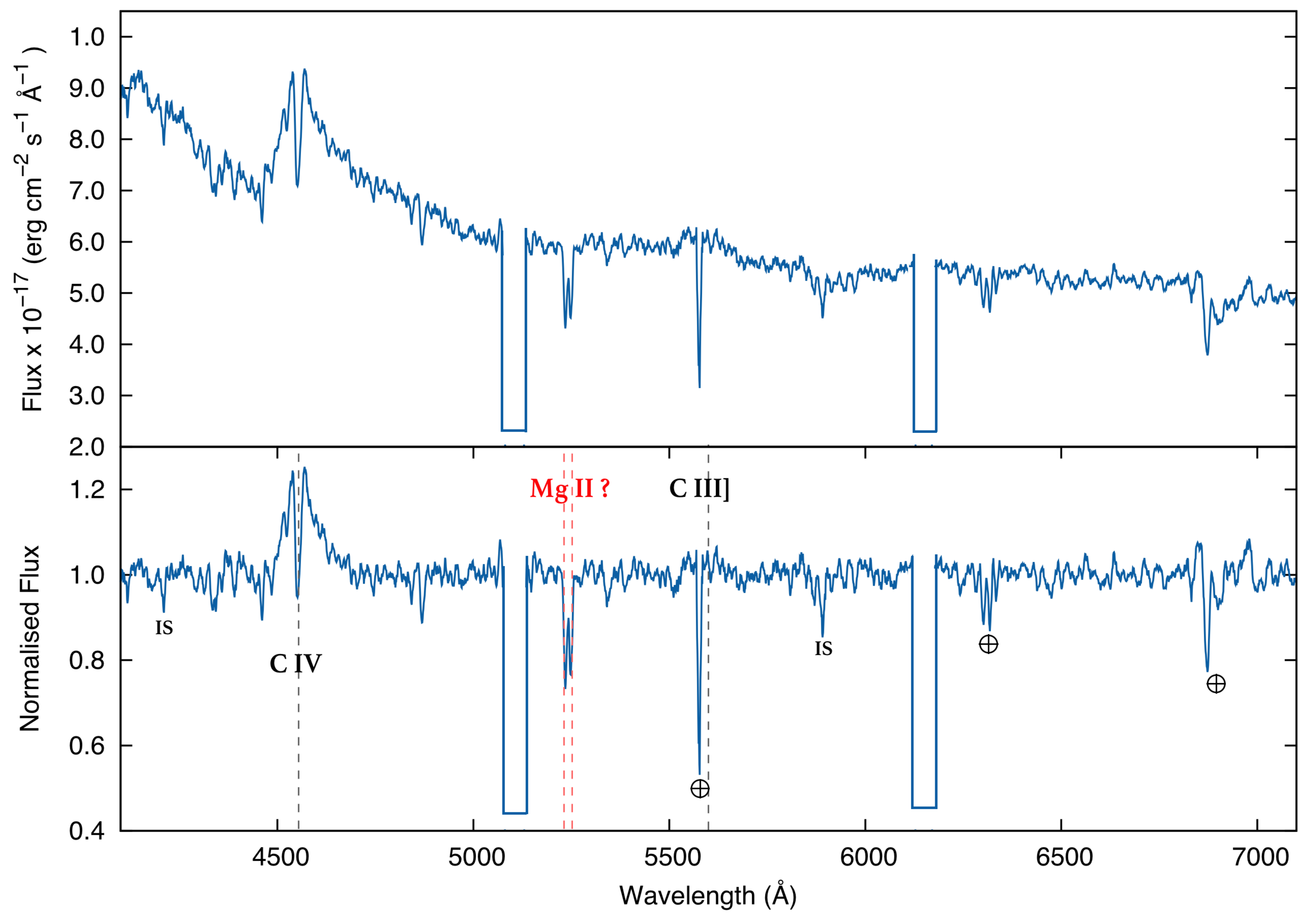}
	\caption{The average spectrum of 3FGL~J0644.3-6713 obtained with RSS/SALT. Two carbon humps features, C~{\sc iv} and C~{\sc iii}] are present. A strong C~{\sc iv} absorption feature is also detected which we suggest is due to material lying further away from the SMBH while the double absorption feature ($\lambda=5240$~\AA{}) we suggest is forbidden Mg~{\sc ii} from an intervening galaxy.}
	\label{fig:J0644}
\end{figure}

A spectrum was obtained for this source in the C (blue) configuration (Fig.~\ref{fig:J0644}). The averaged spectrum features two carbon humps, namely, C~{\sc iv} ($\lambda_{0,{\rm vac}} = 1548.19$~\AA{}) and C~{\sc iii}]
at 4540 and 5586~\AA{}, respectively. An absorption feature, which we also associate with C~{\sc iv} is detected at 4540~\AA{}, suggesting that some material must be lying further away from the central source to produce an absorption feature. The  measurement of the C~{\sc iii}] feature is, unfortunately, hampered by the strong sky line at 5577~\AA{}. 

The broad carbon emission lines give a redshift of $z = 1.930\pm0.004$ while the equivalent width ($|W_\lambda| = 24\pm13$~\AA{}) and FWHM (7446.3~km~s$^{-1}$) of the C~{\sc iv} emission line classify 3FGL~J0644.3-6713 as a FSRQ.

In addition to the two emission features associated with this AGN, there is an absorption feature at 5240~\AA{}. We suggest that this may be the Mg~{\sc ii} line of an intervening galaxy at a redshift of $z\approx0.87$. 

\subsection{3FGL J0730.5-6606}

\begin{figure}
 	\includegraphics[width=\columnwidth]{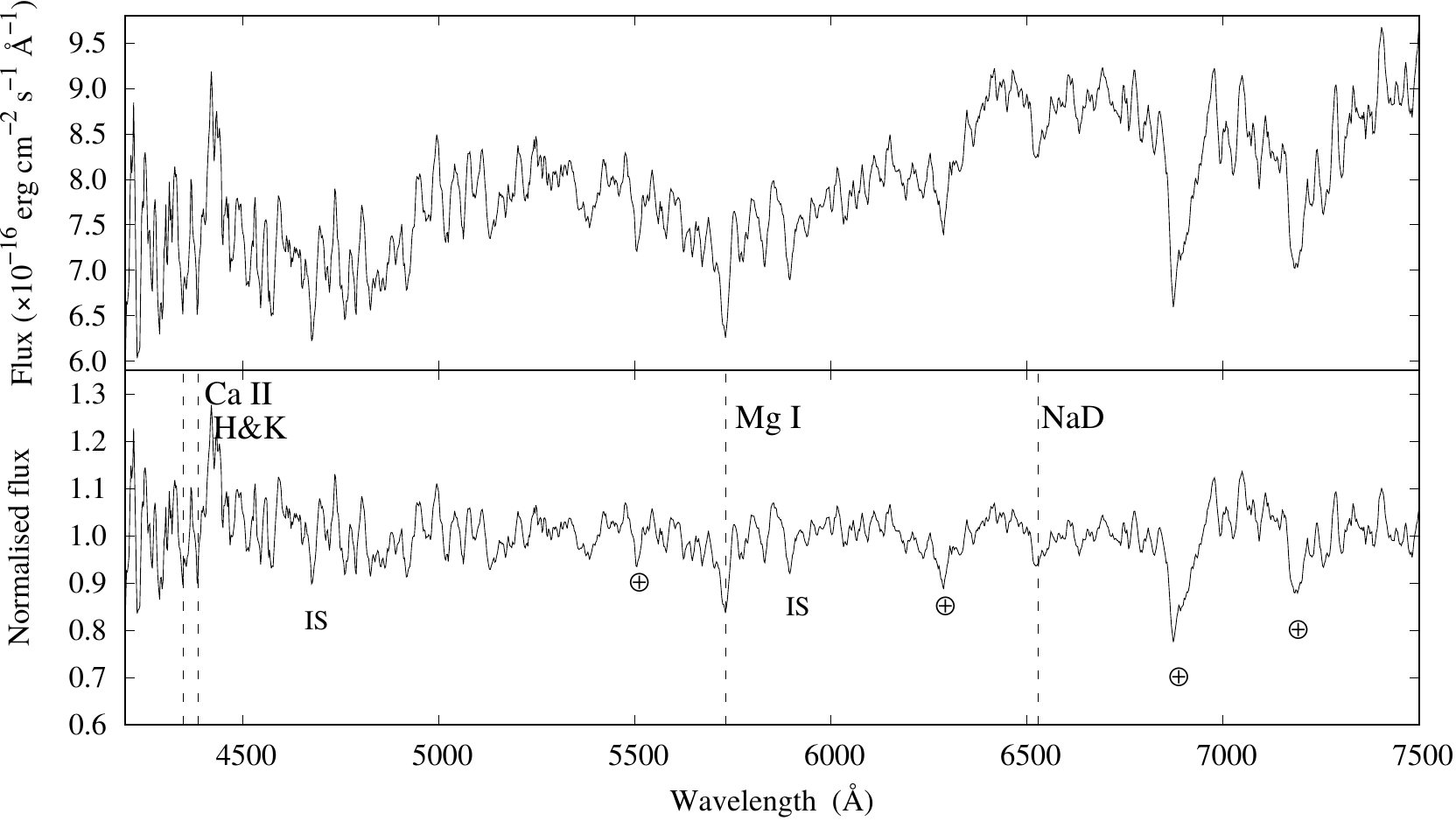}
	\caption{The average spectrum of 3FGL J0730.5-6606 obtained with the SAAO 1.9-m grating spectrograph. Ca~{\sc ii} H\&K, Mg~{\sc i} and NaD absorption lines are detected at an average redshift of $z=0.106 \pm 0.001$.}
	\label{fig:J0730_saao}
\end{figure}

\begin{figure}
 	\includegraphics[width=\columnwidth]{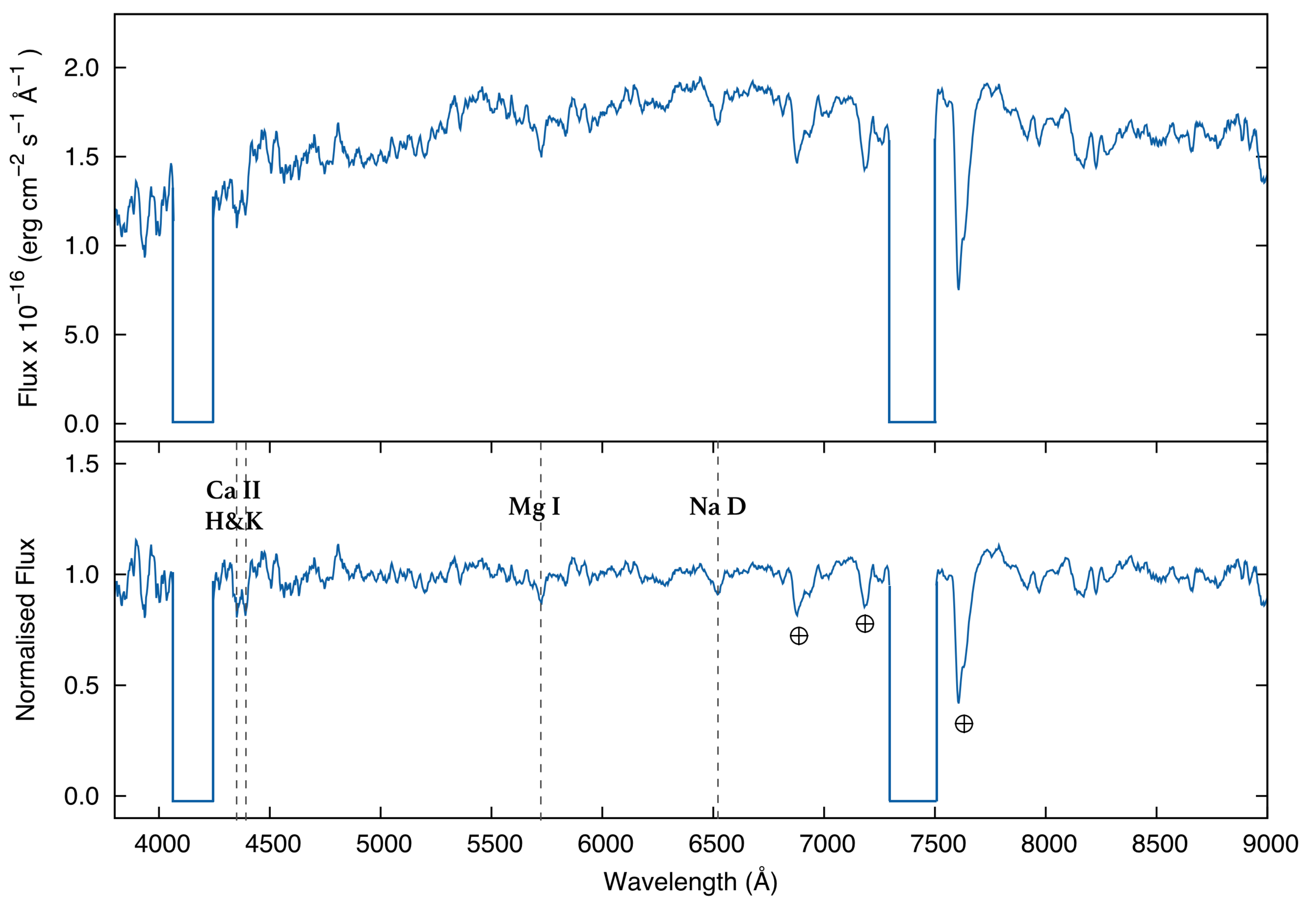}
	\caption{The average spectrum of 3FGL J0730.5-6606 obtained with RSS/SALT with configuration A. Ca~{\sc ii} H\&K, Mg~{\sc i} and NaD absorption lines are detected at an average redshift of $z=0.106 \pm 0.001$.}
	\label{fig:J0730}
\end{figure}

3FGL J0730.5-6606 was observed with the SAAO 1.9-m (Fig.~\ref{fig:J0730_saao}) with follow-up observations with SALT (Configuration A, Fig.~\ref{fig:J0730}). The spectrum shows Ca~{\sc ii} H\&K ($\lambda_\text{0,H} = 3933.66$~\AA{} and $\lambda_\text{0,K} = 3968.47$~\AA{}), Mg~{\sc i} ($\lambda_0 = 5172.68$~\AA{}) and NaD ($\lambda_0 = 5895.92$~\AA{}) absorption lines. The average spectral line position from both the SAAO~1.9-m and SALT observations place the source at a redshift $z=0.106\pm0.001$. 

The AGN nature of the source is clear from the weak Ca break, which we measure as K$_{4000}$ = 0.16. Therefore, 3FGL J0730.5-6606 is classified as a BL Lac.

\subsection{3FGL J1218.8-4827}

\begin{figure}
 	\includegraphics[width=\columnwidth]{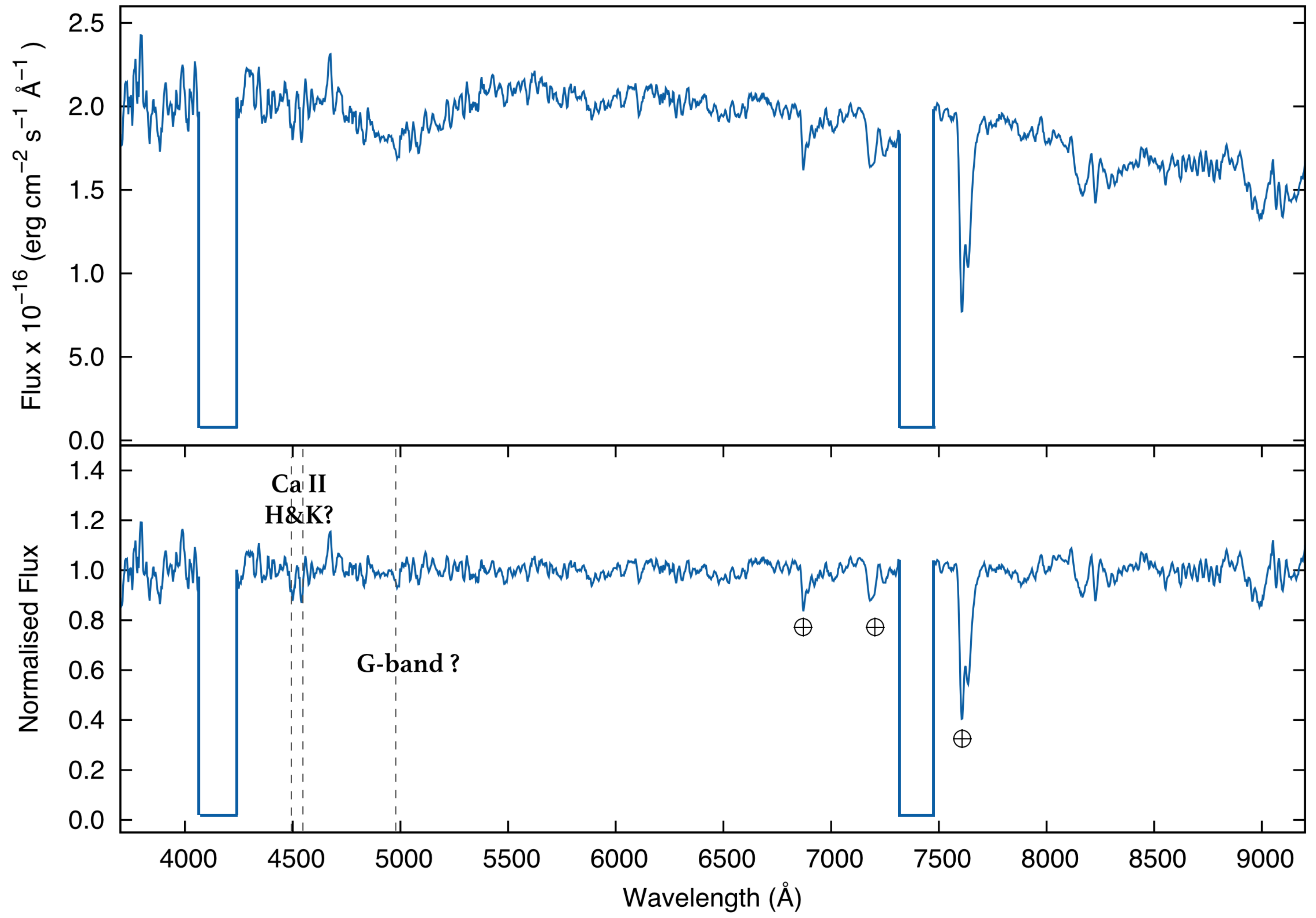}
	\caption{The average spectrum of 3FGL J1218.8-4827 with SALT. Potential Ca~{\sc ii} and G-band absorption lines feature are indicated. }
	\label{fig:J1218}
\end{figure}

The final three sources, all show mainly featureless spectra, with potential spectral features identified. 
SALT spectroscopy of 3FGL J1218.8-4827 (configuration A, S/N~$\sim40$) shows a near featureless spectrum (Fig.~\ref{fig:J1218}), with the indication of potential Ca~{\sc ii} H\&K and G-band absorption features at 4494, 4537 and 4971~\AA{}, respectively. If these features are correct, it places the source at $z=0.150\pm0.006$. Therefore, based on the mainly featureless spectra, we suggest that this is a BL~Lac.

\subsection{3FGL J1407.7-4256}

\begin{figure}
 	\includegraphics[width=\columnwidth]{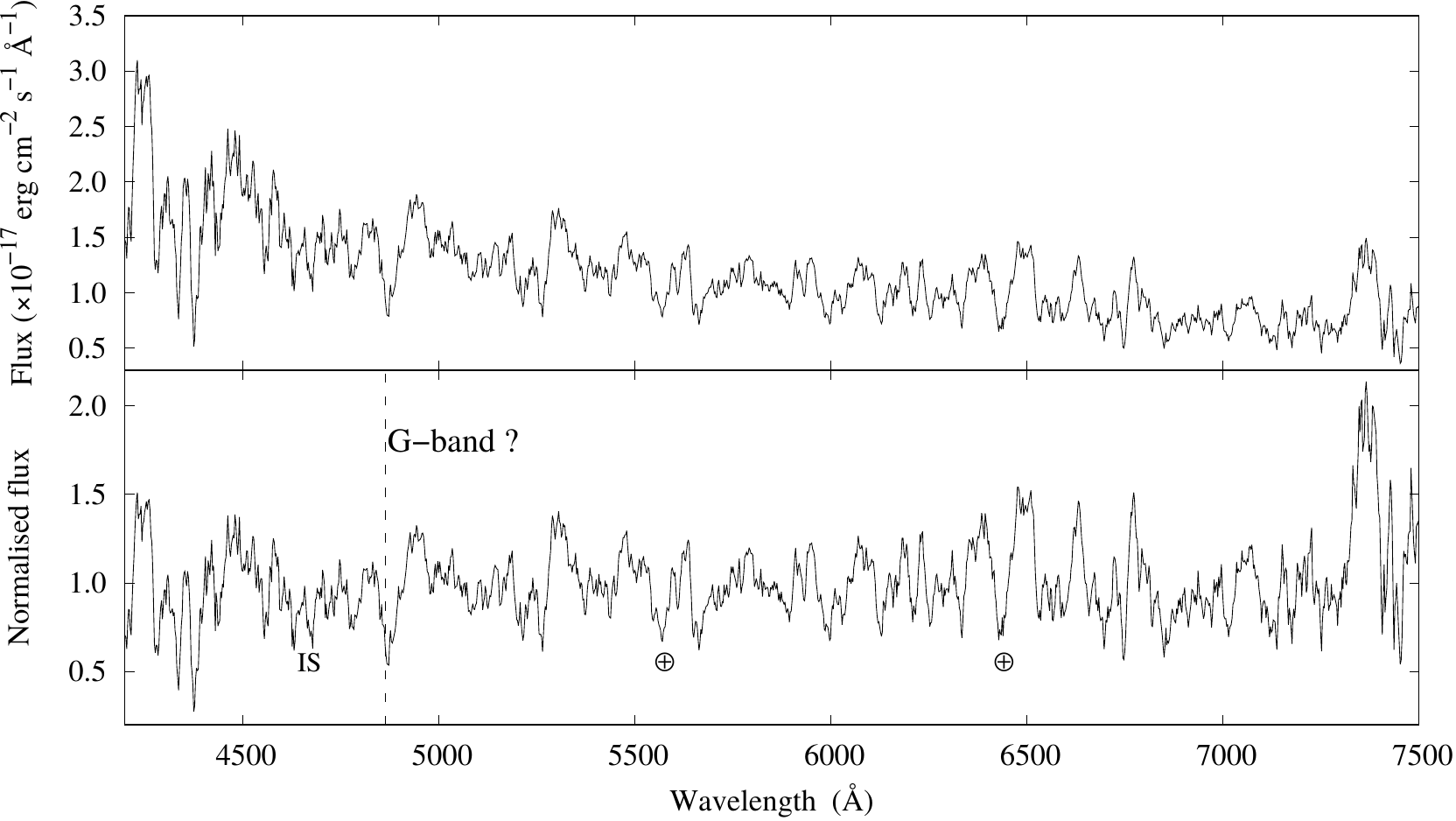}
	\caption{The average 3FGL J1407.7-4256 spectra taken with the SAAO 1.9-m grating spectrograph. The potential G-band lines is tentatively suggested at 4863.63~\AA{}.}	
	\label{fig:J1407_saao}
\end{figure}

\begin{figure}
 	\includegraphics[width=\columnwidth]{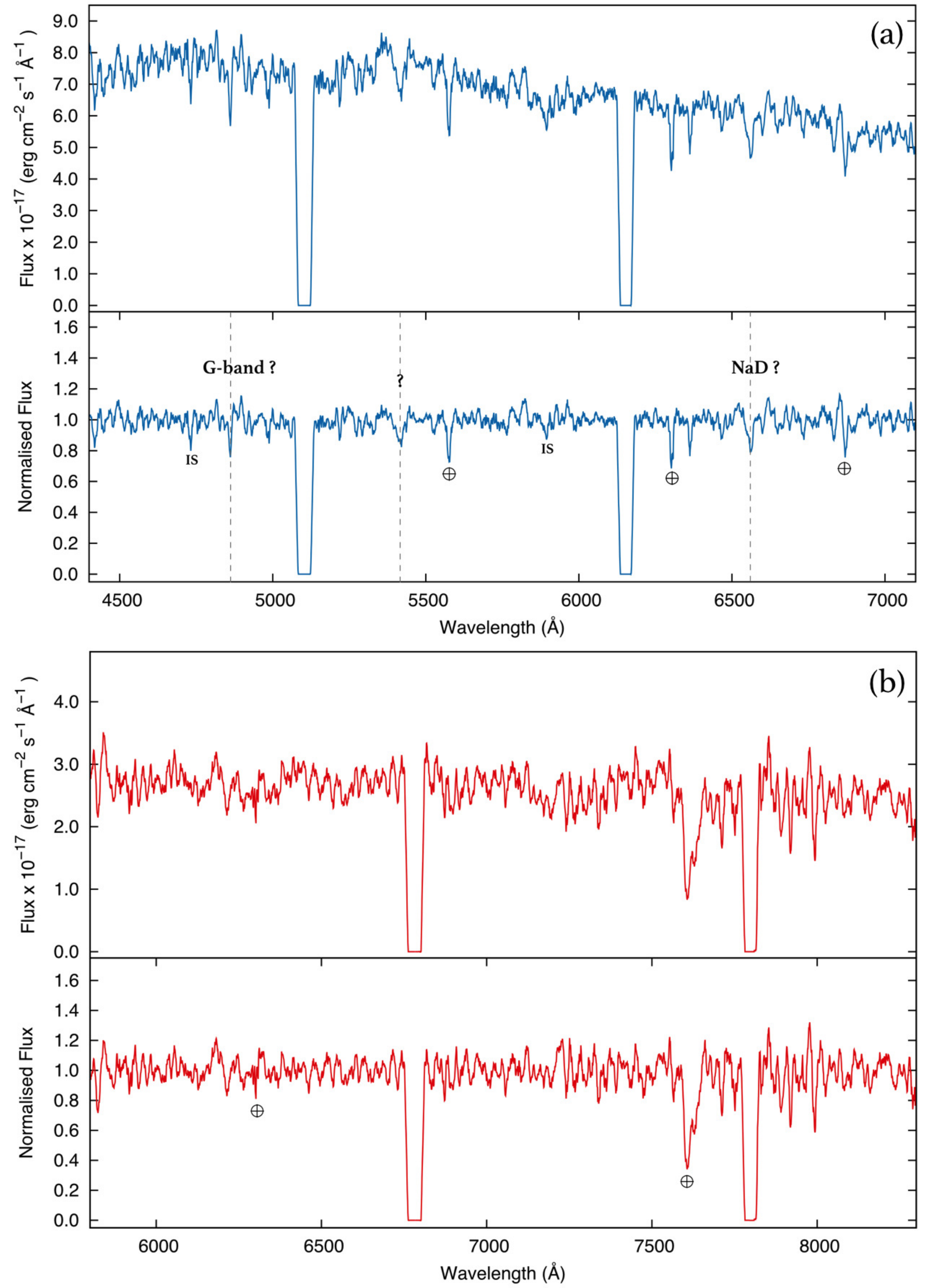}
	\caption{The average 3FGL J1407.7-4256 spectra taken with the RSS for the (a) blue (S/N$\sim$20) and (b) red (S/N$\sim$17) configurations. Potential G-band and NaD absorption lines are tentatively suggested at 4858.79 and 6516.54~\AA{}, respectively, as well as an unknown absorption feature at 5417.07~\AA{}. 
	}
	\label{fig:J1407}
\end{figure}

The second last source, 3FGL J1407.7-4256, was observed with both the SAAO 1.9-m (Fig.~\ref{fig:J1407_saao}) and SALT (C and D configurations, Fig.~\ref{fig:J1407}). However, no clear features are identified.  We, therefore, potentially classify this source as a BL~Lac due to its mainly featureless spectrum and the lack of observed emission lines.

The G-band and NaD absorption lines are tentatively suggested at 4859~\AA{} and 6563~\AA{}, respectively, in the RSS spectrum (Fig.~\ref{fig:J1407}) though follow-up studies should be undertaken to obtain a higher S/N.  If these are correct the source has a redshift of $z=0.124\pm0.009$. 

\subsection{3FGL J2049.7+1002}

\begin{figure}
 	\includegraphics[width=\columnwidth]{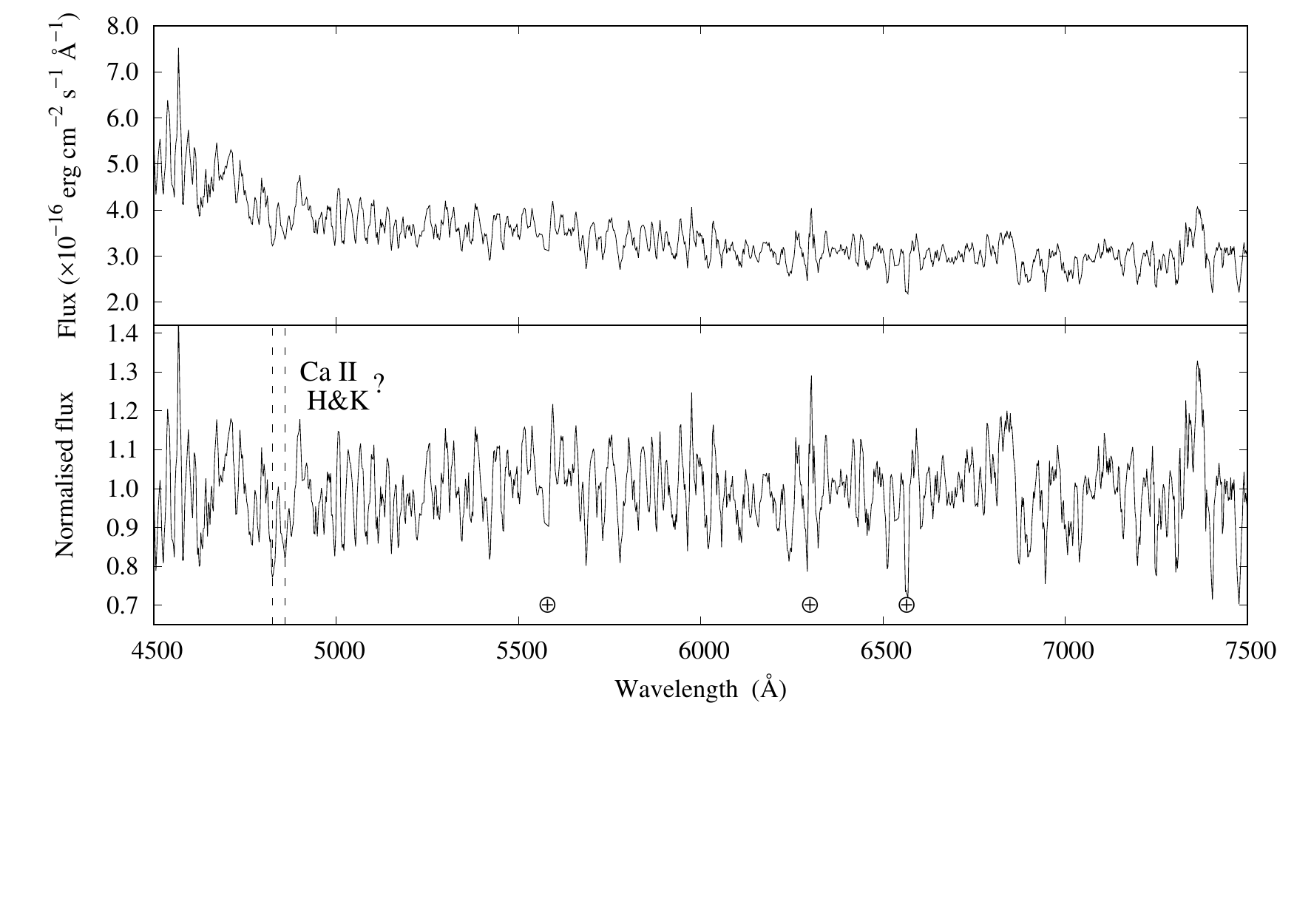}
	\caption{The average 3FGL J2049.7+1002 spectrum taken with the SAAO 1.9-m grating spectrograph. Potential Ca~{\sc ii} H\&K absorption lines lie at 4825 and 4860~\AA{}.}
	\label{fig:J2049}
\end{figure}

3FGL J2049.7+1002 was observed with the SAAO 1.9-m. The averaged spectrum displayed in Fig.~\ref{fig:J2049} yields potential Ca~{\sc ii} H\&K absorption lines at 4824 and 4859~\AA{}. This is, however, tentative given the low signal to noise. If these line identifications are correct the redshift is $z=0.226\pm0.001$. Based on the featureless spectrum we suggest that 2FGL~J2049.8+1001 is a BL~Lac.

\begin{figure}
	% To include a figure from a file named example.*
	% Allowable file formats are eps or ps if compiling using latex
	% or pdf, png, jpg if compiling using pdflatex
	\includegraphics[width=\columnwidth]{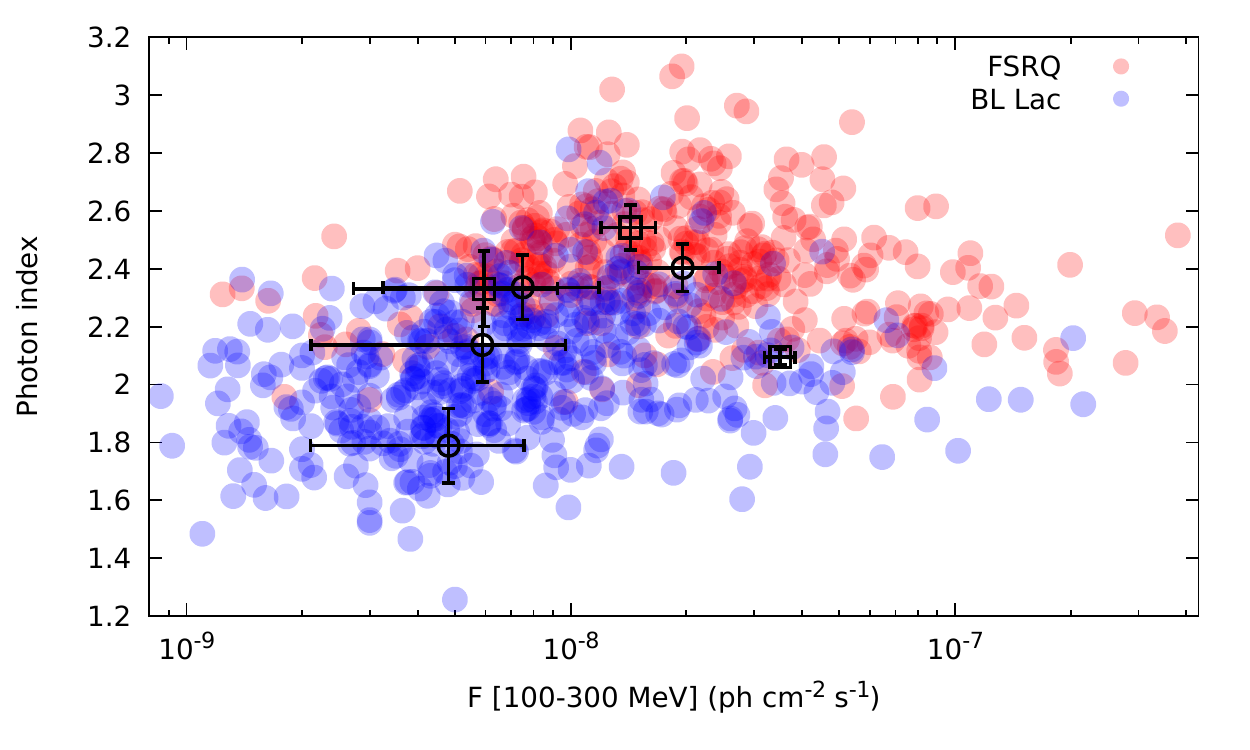}
    \caption{The $\gamma$-ray photon index versus flux of the selected BCUs (black) in comparison to the classified FSRQ (red) and BL Lac (blue) sources. The candidates classified as BL Lacs are marked with open circles and the candidates classified as FRSQs are marked with open squares. The sources all lie within the region expected for blazar sources \citep[see e.g. fig.~8 and discussion in][]{ackermann15}.}
    \label{fig:phot_flux}
\end{figure}

~\\
Fig.~\ref{fig:phot_flux} shows where the selected sources lie in relation to the other identified FSRQs and BL Lacs in the {\it Fermi}-3LAC \citep[see e.g. fig.~8 in][]{ackermann15}. These positions are consistent with the range for blazars.  

\section{Discussion and Conclusions}
\label{sec:conclusion}

Here we have presented the results of optical spectra obtained for seven \fermi\ BCU sources during 2014 and 2015, using the SAAO~1.9-m and RSS on SALT. From the optical spectra we have classified 3 sources as FSRQs and the remaining 4 as BL Lac objects (Table~\ref{tab:source_classification}). These classifications are consistent with the expectation that the selected sources are blazars (Fig.~\ref{fig:phot_flux}). We also note that there is a very good agreement between the results obtained here and the class predictions made by the machine learning techniques by \citet{hassan13} and \citet{chiaro16}, based on their $\gamma$-ray properties.  This is important, since undertaking spectroscopic classification of a large number of sources can be difficult and machine learn can be important for classifying and identifying important sources. Increasing the number of spectroscopically classified sources, as done here, will also provide more sources with which to refine these methods, improving their accuracy. 

Another alternative method for establishing the distance to extra-galactic sources is through photometric redshifts. This has the advantage that it is possible to get higher S/N observations of fainter sources that would not be possible with spectroscopic observations. This technique is more easily applied to large scale surveys \citep[see e.g.][for the 12th SDSS data release]{beck16}. However, photometric redshifts rely on broader wavelength ranges and strong emission features and the determination of redshifts through spectroscopy still achieves a higher accuracy. The spectroscopic observations also allows for a better classification and identification of spectral features which are not measurable through photometric redshifts.

\begin{table}
	\centering
	\caption{Summary of source classification from the optical spectra in this work, with a comparison to the results predicted by the machine learning techniques of \citet{hassan13} and \citet{chiaro16}.}
	\label{tab:source_classification}
	\begin{tabular}{lcccc} % four columns, alignment for each
\hline
3FGL name &   Classification & Hassan et al. & Chiaro et al. \\ \hline
J0045.2-3704 &   FSRQ & FSRQ & FSRQ \\
J0200.9-6635 &   FSRQ & - & - \\
J0644.3-6713 &   FSRQ & - & FSRQ \\
J0730.5-6606 &   BL Lac & BL Lac & BL Lac \\
J1218.8-4827 &   BL Lac & - & - \\
J1407.7-4256 &   BL Lac & BL Lac & BL Lac \\
J2049.7+1002 &   BL Lac & - & - \\ \hline
	\end{tabular}
\end{table}

For 4 of the 7 sources, unambiguous spectral lines have been identified and used to determine redshifts, while for the remaining 3, potential lines have been identified.  The redshift measurements determined for the 7 sources included in our observations are summarised in Table.~\ref{tab:redshifts}. 

For the source, 3FGL~J0730.5-6606, a redshift measurement of $z = 0.1064\pm0.0002$ is given by the 6dF Galaxy Survey \citep{jones09}.  Our results ($z = 0.1063 \pm 0.0009$) confirm this.

\begin{table*}
	\begin{center}
		\caption{\label{tab:redshifts} Spectral lines detected with the Grating Spectrograph/SAAO 1.9-m (SpCCD) and RSS/SALT. Targets marked with an asterisk indicate uncertain line detections and, therefore, uncertain redshifts. The equivalent widths and FWHM are reported for the emission lines.}
		\begin{tabular}{cccccccc}
		  \hline
3FGL name  &   Instrument  & S/N &   Spectral Line  &   $\lambda_\text{obs}$  & $|W_\lambda|$ & FWHM &   $z$    \\ 
  &    &  &    &   (\AA{})  &  & (km~s$^{-1}$) &   \\ \hline 
%  &  &  &  &  &  &  & \\
J0045.2-3704  &   RSS   & 60 &   [C~{ \sc ii}]   & 4730.2 &  $1.5\pm0.8$  & $809\pm14$ &  1.0328 \\ 
  &   RSS   &...  &   Mg~{ \sc ii}   & 5689.2 & $17.9\pm3.4$  &  $2777\pm46$  &  1.0333 \\ \cline{8-8}
  &    &  &    &    &  &  &   $z$ = 1.0331 $ \pm$ 0.0004    \\ \hline
J0200.9-6635  &   RSS   & 25 &   C~{ \sc iii}]   & 4328.42 &  $34\pm13 $ &  $5950\pm240$  &  1.2683 \\ 
  &   RSS   &...  &   Mg~{ \sc ii}   & 6392.29 &  $42\pm12$  &  $2826\pm113$ &  1.2846 \\ 
  &   RSS   & 30 &   Mg~{ \sc ii}   & 6398.72 &  $28.6\pm3.9$  &  $2915\pm88$  &  1.2869 \\ \cline{8-8}
  &    &  &    &    &  &  &   $z$ = 1.28 $ \pm$ 0.01    \\ \hline
J0644.3-6713  &   RSS   & 60 &   C~{ \sc iv}   & 4539.76 &  $24\pm13$  &  $7445\pm124$  &  1.9331 \\ 
  &   RSS   &...  &   C~{ \sc iii}]   & 5586.4 &  &  &  1.9275 \\ \cline{8-8}
  &    &  &    &    &  &  &   $z$ =  1.930 $ \pm$ 0.004     \\ \hline
J0730.5-6606  &   SpCCD   & 15 &   Ca~{ \sc ii} H   & 4349.05 &  &  &  0.1056 \\ 
  &   SpCCD   &...  &   Ca~{ \sc ii} K   & 4387.14 &  &  &  0.1055 \\ 
  &   SpCCD   &...  &   Mg~{ \sc i}   & 5730.81 &  &  &  0.1079 \\ 
  &   SpCCD   &...  &   NaD   & 6529.73 &  &  &  0.1075 \\ 
  &   RSS   & 27 &   Ca~{ \sc ii} H   & 4349.06 &  &  &  0.1056 \\ 
  &   RSS   &...  &   Ca~{ \sc ii} K   & 4388.94 &  &  &  0.1057 \\ 
  &   RSS   & ... &   Mg~{ \sc i}   & 5721.5 &  &  &  0.1061 \\ 
  &   RSS   &..  &   NaD   & 6524.43 &  &  &  0.1066 \\ \cline{8-8}
  &    &  &    &    &  &  &   $z$ = 0.106 $ \pm$ 0.001     \\ \hline
J1218.8-4827*  &   RSS   & 42 &   Ca~{ \sc ii} H ?   & 4494.21 &  &  &  0.1425 \\ 
  &   RSS   & ... &   Ca~{ \sc ii} K ?   & 4537.12 &  &  &  0.152 \\ 
  &   RSS   & ... &   G-band ?   & 4970.85 &  &  &  0.1544 \\ \cline{8-8}
  &    &  &    &    &  &  &   $z$ = 0.150 $ \pm$ 0.006     \\ \hline
J1407.7-4256*  &   SpCCD   & 10 &   G-band ?   & 4863.63 &  &  &  0.1295 \\ 
  &   RSS   & 20 &   G-band ?   & 4858.89 &  &  &  0.1284 \\ 
  &   RSS   & ... &   NaD ?   & 6563.93 &  &  &  0.1133 \\ \cline{8-8}
  &    &  &    &    &  &  &   $z$ = 0.124 $ \pm$ 0.009     \\ \hline
J2049.7+1002*  &   SpCCD   & 15 &   Ca~{ \sc ii} H ?   & 4824.45 &  &  &  0.2262 \\ 
  &   SpCCD   &...  &   Ca~{ \sc ii} K ?   & 4858.6 &  &  &  0.2243 \\ \cline{8-8}
    &    &  &    &    &  &  &   $z$ = 0.226 $ \pm$ 0.001     \\ \hline
	  \end{tabular}

	\end{center}
\end{table*}

Finally we note that 3FGL~J0730.5-6606 is also the best candidate identified from our sample for VHE observations since it lies at a low redshift and is detected above 50~GeV as listed in the \fermi\ catalogue of hard $\gamma$-ray sources \citep[2FHL,][]{ackermann16}.

As mentioned all the sources discussed here were originally selected from the 2LAC catalogue, however, a newer release (3LAC) was subsequently published, which has greatly increased the number of detected $\gamma$-ray sources, including unclassified blazar candidates and unassociated, sources. In addition, 60 of the sources identified as potential TeV sources in the 2FHL catalogue are listed as BCUs (one of which, 3FGL~J0730.5-6606, we have classified as a BL Lac). Given the improved sensitivity at VHE that will be possible with the CTA \citep{actis11} it is important to continue to classify these sources and search for new VHE candidates. Therefore, our observations of BCU sources are continuing with new candidates selected from these catalogues.

\section*{Acknowledgements}

Observations reported in this paper were obtained with the Southern African Large Telescope (SALT) under programs 2014-2-SCI-055 and 2015-1-SCI-053. This paper uses observations made at the South African Astronomical Observatory (SAAO). BvS acknowledges that this work is based on the research supported in part by the National Research Foundation (NRF) of South Africa for the grant, No. 87919. PV also acknowledges support from the NRF of SA.

%%%%%%%%%%%%%%%%%%%%%%%%%%%%%%%%%%%%%%%%%%%%%%%%%%

%%%%%%%%%%%%%%%%%%%% REFERENCES %%%%%%%%%%%%%%%%%%

% The best way to enter references is to use BibTeX:

\bibliographystyle{mnras}
\bibliography{bcu_ref} % if your bibtex file is called example.bib

% Alternatively you could enter them by hand, like this:
% This method is tedious and prone to error if you have lots of references
% \begin{thebibliography}{99}
% \bibitem[\protect\citeauthoryear{Author}{2012}]{Author2012}
% Author A.~N., 2013, Journal of Improbable Astronomy, 1, 1
% \bibitem[\protect\citeauthoryear{Others}{2013}]{Others2013}
% Others S., 2012, Journal of Interesting Stuff, 17, 198
% 

% \bibitem[\protect\citeauthoryear{]{}
% 
% Urry C M and Padovani P 1995 PASP 107 803
% 

% \end{thebibliography}

%%%%%%%%%%%%%%%%%%%%%%%%%%%%%%%%%%%%%%%%%%%%%%%%%%

%%%%%%%%%%%%%%%%% APPENDICES %%%%%%%%%%%%%%%%%%%%%

% \appendix
% 
% \section{Some extra material}
% 
% If you want to present additional material which would interrupt the flow of the main paper,
% it can be placed in an Appendix which appears after the list of references.

%%%%%%%%%%%%%%%%%%%%%%%%%%%%%%%%%%%%%%%%%%%%%%%%%%

% Don't change these lines
\bsp	% typesetting comment
\label{lastpage}
\end{document}